\documentclass[12pt]{article}
\usepackage{graphicx}

\begin{document}


\section*{How can we distinguish positive cooperativity from auto-catalysis in enzyme kinetics?}

Sharmistha Dhatt$\,^1$, Kinshuk Banerjee$\,^2$ and 
Kamal Bhattacharyya$\,^{1,*}$  

\noindent
$\,^1$Dept. of Chemistry, University of Calcutta, 
92 A.P.C. Road, Kolkata 700 009, India. \\
$\,^2$Dept. of Chemistry, A.J.C. Bose College, 
1/1B A.J.C. Bose Road, Kolkata 700 020. $\,^*$ E-mail: pchemkb@gmail.com

\begin{abstract}
Different graphical plots involving the catalytic rate with 
the (initial) substrate concentration exist in the enzyme kinetics literature to estimate the reaction constants. But, none of these standard plots can 
unambiguously distinguish between the two important mechanisms of rate enhancement: positive 
cooperativity among the active sites of an oligomeric enzyme and 
auto-catalysis of the intermediate complex of an enzyme with a 
single active site. 
We achieve this distinction here by providing a nice linear plot for the latter. Importantly, to accomplish this task, no extra information 
other than the steady-state rate as a function of substrate 
concentration is required. 

\end{abstract}

Keywords: Enzyme kinetics; Cooperativity; Auto-catalysis; Steady state


\section{Introduction}

Enzyme catalysis is a highly important biochemical reaction where specific substrates are efficiently converted into products \cite{mar}. Investigations on the catalytic mechanisms span over a century. 
The Michaelis-Menten (MM) scheme \cite{mm} is a cornerstone in the field of 
theoretical modelling of enzyme kinetic data \cite{cor}. The rate of product formation in MM scheme under steady-state approximation 
\cite{seg} of the intermediate complex plays the benchmark role in the analyses of kinetic constants. 
To determine the rate parameters, various ways of plotting the rate against (initial) substrate concentration are present in the literature with their respective advantages and disadvantages \cite{cor}. 
Some notable examples are the Lineweaver-Burk (LB), Eadie-Hofstee (EH) 
and Hanes-Woolf (HW) plots \cite{mar,cor}.

The MM scheme represents the simplest model of enzyme catalysis. Naturally, the enzyme is considered to have a single substrate binding site or active site. 
However, there are many enzymes in nature with multiple binding sites. Interactions among these sites leads to cooperativity \cite{Ham}. 
As a result, one notices either an enhancement (positive cooperativity) 
or diminution (negative cooperativity) of the catalytic rate compared to that obtained from the MM scheme with equal number of binding sites acting independently, {\it i.e.}, zero cooperativity \cite{Ban}. 
A prime example of cooperative kinetics is the oxygen binding 
to hemoglobin \cite{Palm}. 
Another possible way of rate enhancement is auto-catalysis \cite{Ostw} 
of the intermediate complex. This can also increase the rate compared 
to the MM case 
at similar substrate concentration even for an enzyme with a 
single active site. Auto-catalysis is thought to play crucial roles in 
the evolution of population \cite{lot} as well as gene \cite{mul}. 
A well-known signature of MM kinetics 
is the hyperbolic curve of rate against starting substrate concentration, 
$[S]_0$. 
Now, both positive cooperativity and auto-catalysis can show non-hyperbolic sigmoidal nature \cite{plas} of the rate against $[S]_0$. Thus, these 
rate-enhancing mechanisms can be identified, in principle, as non-MM ones. 
Then, the important question to ask is: 
How to distinguish 
between auto-catalysis and positive cooperativity from catalytic rate measurements?

In this work, we show that none of the standard plots can unambiguously discriminate auto-catalytic behaviour from positive cooperativity. As a remedy, we introduce here a graphical method that can do the job quite well. In what follows, cooperativity stands for positive cooperativity only.

\begin{figure}[tbh]
\centering
\rotatebox{0}{
\includegraphics[width=7cm,keepaspectratio]{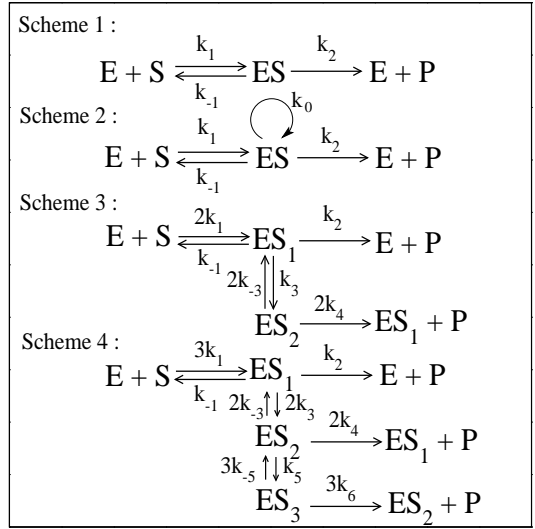}}
\caption{Reaction schemes for the various cases considered. MM: Michaelis-Menten; Auto: auto-catalysis (of the intermediate); 2-site: cooperative 
kinetics for an enzyme with two active sites; 3-site: cooperative 
kinetics for an enzyme with three active sites. The factors `2' and 
`3' account for the statistical weightage.}
\label{fig1}
\end{figure}

\section{Steady-state catalytic rates}

We determine the rates of product formation in all the cases 
under the steady-state approximation (SSA) of the respective 
intermediate complexes. The calculations are based on the schemes 
shown in Fig. \ref{fig1}. We take $[E]_0=1\,\mu M$ and 
$[S]_0\ge 100\,\mu M$. The validity of SSA is checked for two 
different choices of the set of rate constants. 
This also ensures $\overline{[S]}\sim [S]_0$ in all the cases. 

\subsection{MM kinetics}

The steady-state (SS) rate of product formation comes out as
\begin{equation}
\overline{R}=\frac{k_2 [E]_0 \overline{[S]}}{K_{\rm M}+\overline{[S]}}
\label{rMM}
\end{equation}
where $[X]_0$ is the starting concentration of the species $X$ and 
$\overline{[X]}$ denotes SS concentration of $X$. 
The MM constant is $K_{\rm M}=\frac{k_{-1}+k_2}{k_1}$. 
The conservation of enzyme and substrate concentrations are given by
\begin{equation}
[E]_0=[E]+[ES]\,;\,[S]_0=[S]+[ES]+[P].
\label{con1}
\end{equation}

\subsection{Auto-catalysis}

In this case, the SS rate is given as
\begin{equation}
\overline{R}=\frac{k_2 [E]_0\overline{[S]}}
{K_{\rm M}+\overline{[S]}-\frac{k_0\overline{[E]}\, \overline{[S]}}{k_1}}.
\label{rauto}
\end{equation}
The conservation conditions are the same as given in Eq.(\ref{con1}).

\subsection{2-site cooperativity}

The expression of SS rate for this scheme is
\begin{equation}
\overline{R}=\frac{2k_4 [E]_0\overline{[S]}}{K_{\rm M}K_{\rm M1}+2K_{\rm M1}\overline{[S]}+\overline{[S]}^2}
\left(k_2K_{\rm M1}/k_4+\overline{[S]}\right).
\label{r2site}
\end{equation}
Here, $K_{\rm M1}=\frac{k_{-3}+k_4}{k_3}$ and $K_{\rm M}$ is as defined above for the MM case.  
The conservation relations used are
\begin{equation}
[E]_0=[E]+[ES]_1+[ES]_2\,;\, [S]_0=[S]+[ES]_1+2[ES]_2+[P].
\label{con3}
\end{equation}

\subsection{3-site cooperativity}

This scheme has a SS rate of
$$
\overline{R}=\frac{3k_6 [E]_0\overline{[S]}}
{K_{\rm M}K_{\rm M1}K_{\rm M2}+
3K_{\rm M1}K_{\rm M2}\overline{[S]}+3K_{\rm M2}\overline{[S]}^2+
\overline{[S]}^3}\, \times$$
\begin{equation}
\left(\frac{k_2 K_{\rm M1}K_{\rm M2}}{k_6}+
\frac{2k_4 K_{\rm M2}\overline{[S]}}{k_6}+
\overline{[S]}^2\right).
\end{equation}
Here, $K_{\rm M2}=\frac{k_{-5}+k_6}{k_5}$ and $K_{\rm M},\,K_{\rm M1}$ 
are as defined already. 
The conservation relations are obtained as natural extensions of 
Eq.(\ref{con3})
\begin{equation}
[E]_0=[E]+[ES]_1+[ES]_2+[ES]_3\,;\,
[S]_0=[S]+[ES]_1+2[ES]_2+3[ES]_3+[P].
\end{equation}

\section{Non-MM behavior of cooperative and auto-catalytic kinetics}

\begin{figure}[tbh]
\centering
\rotatebox{0}{
\includegraphics[width=9cm,keepaspectratio]{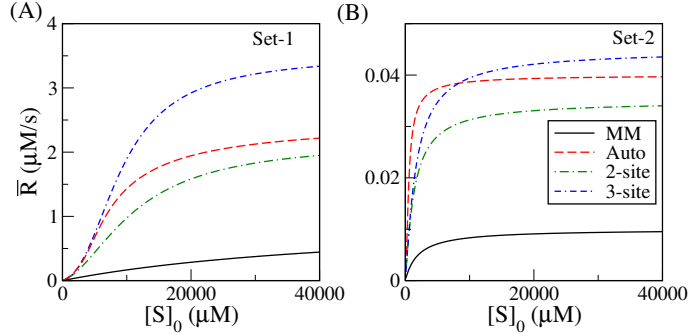}}
\caption{Variation of the SS catalytic rate $\overline{R}$ 
as a function of the starting 
substrate concentration $[S]_0$ for the different kinetic schemes 
with two different sets of rate constants plotted in (A) and (B).  }
\label{fig2}
\end{figure}

The nature of variation of the SS catalytic rate, $\overline{R}$, 
with $[S]_0$ can give a nice indication of non-MM behavior. The 
curve is hyperbolic for MM kinetics but generally sigmoidal for 
cooperativity and auto-catalysis. This is evident from 
Fig. \ref{fig2} for two different choices of rate constants 
(see Table \ref{tab1}). 
The two sets of rate constants can result in 
up to 100-times difference in the respective SS rates. 
It is clear from Fig. \ref{fig2} that cooperativity and auto-catalysis can not be 
distinguished from each other, as already mentioned in Section 1. 

\begin{table}[tbh]
\label{tab1}
\caption{Rate constants for the various cases considered in the main text. $k_1,\,k_3,\,k_5$ are in $(\mu M.s)^{-1}$, $k_0$ is in 
$\mu M^{-2}.s^{-1}$, $k_{-1},\,k_2,\,k_{-3},\,k_4,
\,k_{-5},\,k_6$ are in $s^{-1}$.}
\begin{tabular}{c|c|c|c|c|c|c|c|c|c|c|c}
\hline
\hline
Set & Case & $k_1$ & $k_{-1}$ & $k_2$ & $k_0$ & 
$k_3$ & $k_{-3}$ & $k_4$ & $k_5$ & $k_{-5}$ & $k_6$\\
\hline
   & MM & 0.1 & 200 & 0.01 & - & - & - & - & - & - & -\\
1 & Auto & 0.01 & 500 & 2.5 & 0.1 & - & - & - & - & -& - \\
  & 2-site & 0.008 & 500 & 1.0 & - & 0.05 & 100 & 1.1 & - & - &-\\
  & 3-site & 0.006 & 500 & 1.0 & - & 0.02 & 100 & 1.1 & 0.05 & 100 & 1.2\\
\hline
  & MM & 0.1 & 200 & 0.01 & - & - & - & - & - & - & -\\
2 & Auto & 0.1 & 200 & 0.04 & 0.5 & - & - & - & - & -& - \\
  & 2-site & 0.2 & 250 & 0.01 & - & 0.25 & 200 & 0.0175 & - & - & -\\
  & 3-site & 0.2 & 350 & 0.01 & - & 0.25 & 325 & 0.0125 & 0.3 & 
300 & 0.015\\
\hline
\end{tabular}
\end{table}

\begin{figure}[tbh]
\centering
\rotatebox{0}{
\includegraphics[width=9cm,keepaspectratio]{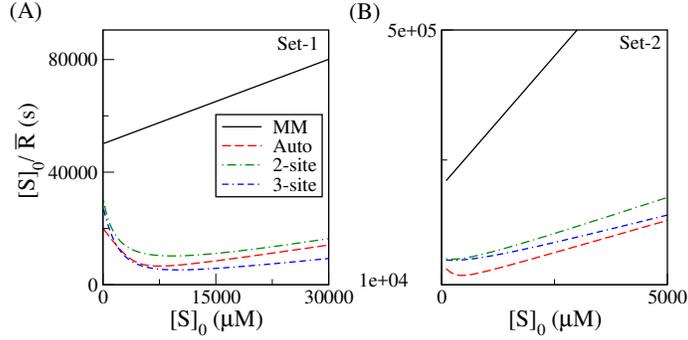}}
\caption{Hanes-Woolf plots for the various schemes with two different sets of rate constants shown in (A) and (B). }
\label{fig3}
\end{figure}

One important point to mention is that, if the degree of cooperativity 
or auto-catalysis is small, the sigmoidal nature of the curve 
may be difficult to identify. Hence, the curves for all the cases 
may appear similar, {\it i.e.}, hyperbolic. In such cases, and also 
generally, plots like HW and LB may help to 
distinguish the non-MM behavior. For example, in Fig. \ref{fig3}, 
we show the HW plots for all the cases taking the two 
sets of rate constants. The MM case has a positive slope throughout 
whereas, rest of the schemes have negative slopes at lower 
range of $[S]_0$. 
But, even these plots can not unambiguously distinguish cooperativity from auto-catalysis. Similar observations are made with the LB and EH plots.

\section{Method to discriminate auto-catalysis from cooperativity}

\begin{figure}[tbh]
\centering
\rotatebox{0}{
\includegraphics[width=9cm,keepaspectratio]{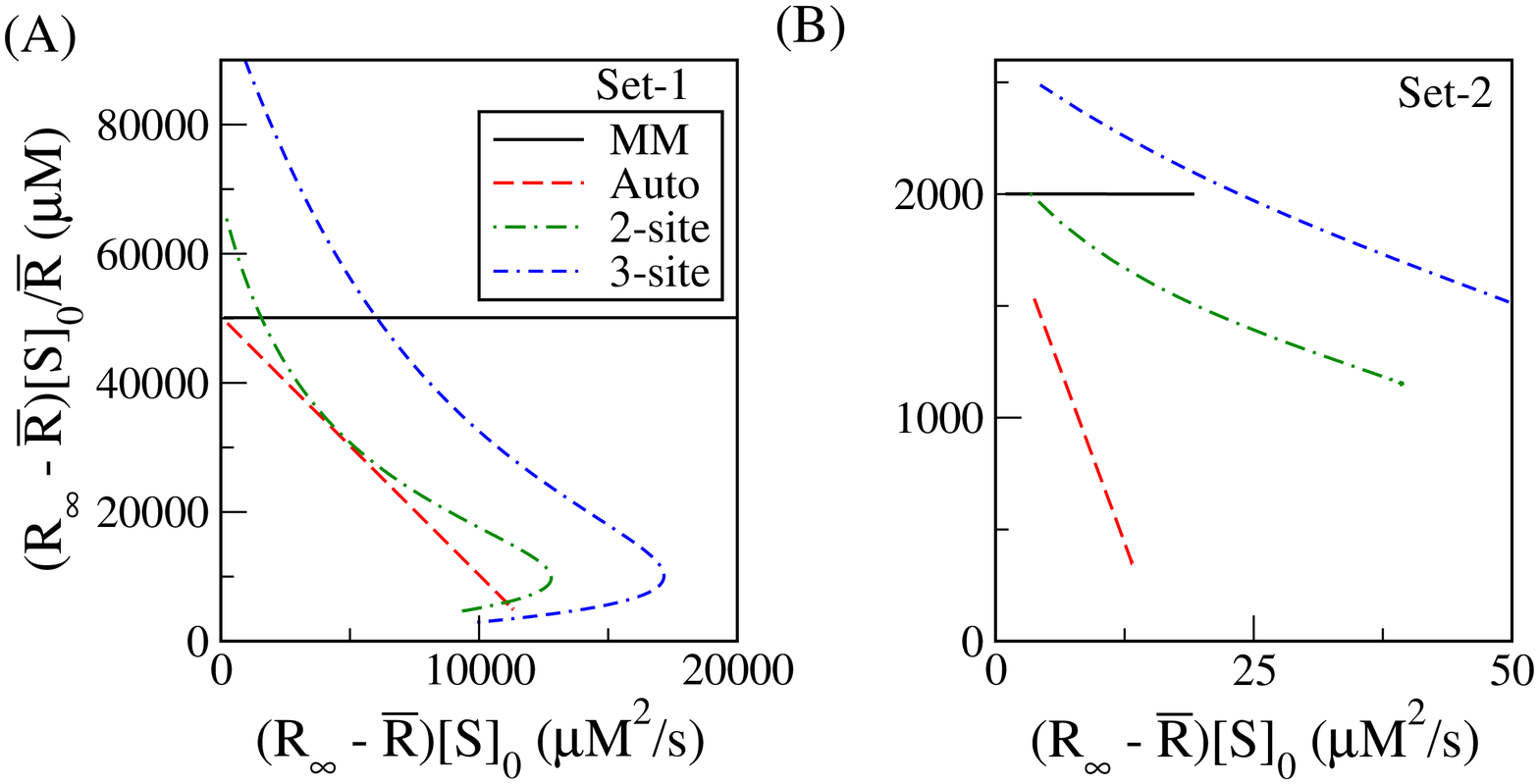}}
\caption{ Plot of $(R_{\infty}-\overline{R})\overline{[S]}/\overline{R}$ against $(R_{\infty}-\overline{R})\overline{[S]}$ for all the 
schemes with the two different sets of rate constants shown in (A) and (B).}
\label{fig4}
\end{figure}

The expression of rate under SSA for the auto-catalysis scheme can be rearranged to the form 
\begin{equation}
(R_{\infty}-\overline{R})\overline{[S]}/\overline{R}=
K_{\rm M}-\frac{k_0}{k_1k_2}(R_{\infty}-\overline{R})\overline{[S]}.
\label{plot}
\end{equation}
Here $R_{\infty}$ is the saturated catalytic rate obtained at large 
substrate concentration. In case of auto-catalysis, it is given by
$R_{\infty}=k_2 [E]_0$. It follows from Eq.(\ref{plot}) that a plot 
of $(R_{\infty}-\overline{R})\overline{[S]}/\overline{R}$ against 
$(R_{\infty}-\overline{R})\overline{[S]}$ yields a straight line with 
negative slope. 
This is clearly shown in Fig. \ref{fig4}.
However, for the cooperative kinetics, the plots 
become non-linear. For example, for the 2-site cooperative scheme, 
we can write
\begin{equation}
(R_{\infty}-\overline{R})\overline{[S]}/\overline{R}=
f(\overline{[S]})(R_{\infty}-\overline{R})\overline{[S]}.
\label{plot1}
\end{equation}
Here, 
\begin{equation}
f(\overline{[S]})=
\frac{K_{\rm M}K_{\rm M1}+2K_{\rm M1}\overline{[S]}+\overline{[S]}^2}
{R_{\infty}\overline{[S]}}
\left(\frac{k_2K_{\rm M1}}{k_4}+\overline{[S]}\right)^{-1}
\end{equation}
and $R_{\infty}=2k_4[E]_0$.
The non-linear nature of the plot is depicted in Fig. \ref{fig4}. 
As the substrate is in excess, we replace $\overline{[S]}$ 
by $[S]_0$ without introducing any significant error. 
For 3-site cooperativity, $R_{\infty}=3k_6[E]_0$.

\section{Conclusion}

The standard graphical methods to characterize the nature of enzyme 
kinetics fail to distinguish auto-catalysis of the intermediate 
in an enzyme with a single active site from positive cooperativity among active sites in an oligomeric enzyme. 
Here, we introduce an approach to attain this discrimination between 
these rate-enhancing mechanisms. 
Our method should be accurate provided (i) the SSA holds for the 
intermediate(s) and (ii) $R_{\infty}$ can be measured with reasonable 
accuracy. Condition (i) is expected to be maintained for any 
standard laboratory experiment and consequent modelling of enzyme kinetics. In fact, all the standard plots in literature are 
based on the validity of SSA. Condition (ii) also can be fulfilled 
either using direct 
measurement or extrapolation of rate data. Thus, we believe that, 
the graphical approach introduced here should be convenient 
to get a clear distinction between auto-catalysis and positive cooperativity.



\begin{thebibliography}{99}

\bibitem{mar} A. G. Marangoni, {\it Enzyme Kinetics: 
A Modern Approach} (Wiley-Interscience, NJ, 2003).

\bibitem{mm} L. Michaelis and M. L. Menten, Biochem. Z. {\bf 49}, 
333 (1913); K. A. Johnson and R. S. Goody, 
Biochemistry, {\bf 50}, 8264 (2011).

\bibitem{cor} A. Cornish-Bowden, {\it Fundamentals of Enzyme 
Kinetics}, 4th edn. (Wiley-VCH, Weinheim, 2012).

\bibitem{seg} L. A. Segel and M. Slemrod, SIAM Review {\bf 31}, 446 
(1989).

\bibitem{Ham} G. Hammes and C. W. Wu, Annu. Rev. Biophys. Bioeng. 
{\bf 3}, 1 (1974).

\bibitem{Ban} K. Banerjee, B. Das, G. Gangopadhyay, 
J. Chem. Phys. {\bf 136}, 154502 (2012).

\bibitem{Palm} T. Palmer and P. Bonner, {\it Enzymes: Biochemistry, Biotechnology, and Clinical
Chemistry}, 2nd ed. (Horwood, West Sussex, 2007).

\bibitem{Ostw} W. Ostwald, {\it In Outlines of General Chemistry}, 
(Macmillan and Co., London, 1912).

\bibitem{lot} A. J. Lotka, Proc. Natl. Acad. Sci. U.S.A. {\bf 6}, 410 (1920).
\bibitem{mul} H. J. Muller, Am. Nat. {\bf 56}, 32 (1922).
\bibitem{plas} R. Plasson, A. Brandenburg, L. Jullien, H. Bersini, 
J. Phys. Chem. A, {\bf 115}, 8073 (2011).

\end{thebibliography}
\end{document}